%% file: main.tex
\documentclass[sigconf]{acmart}

\copyrightyear{2026}
\acmYear{2026}
\setcopyright{cc}
\setcctype{by}
\acmConference[UbiComp Companion '26]{Companion of the 2026 ACM International Joint Conference on Pervasive and Ubiquitous Computing}{October 11--15, 2026}{Shanghai, China}
\acmBooktitle{Companion of the 2026 ACM International Joint Conference on Pervasive and Ubiquitous Computing (UbiComp Companion '26), October 11--15, 2026, Shanghai, China}
\acmDOI{10.1145/3798063.3837278}
\acmISBN{979-8-4007-2533-3/2026/10}

\newcommand{\systemname}{\textit{Calmables}}
\newcommand{\systemnameheading}{Calmables}

\usepackage{todonotes}
\usepackage{balance}

\newbool{showComments}
\booltrue{showComments}
\usepackage{listings}
\usepackage{array}
\usepackage{multirow}
\usepackage{xcolor}
\usepackage{etoolbox} 
\usepackage{soul}
\ifbool{showComments}{%
\newcommand{\tr}[1]{\sethlcolor{blue!20}\hl{[Tobi: #1]}}
}{%
  \newcommand{\tr}[1]{}%
}

\input{statistics.tex}

\begin{document}

%%
%% The "title" command has an optional parameter,
%% allowing the author to define a "short title" to be used in page headers.
% \title{UltrasonicSpeaker: Localized Multi-Language Audio Delivery via Ultrasound and Wearable Earphones}
\title{\systemnameheading{}: Demonstrating Closed-Loop Infrared Earables for Thermal Biofeedback and Relaxation Support}

%%
%% The "author" command and its associated commands are used to define
%% the authors and their affiliations.
%% Of note is the shared affiliation of the first two authors, and the
%% "authornote" and "authornotemark" commands
%% used to denote shared contribution to the research.
\author{Valeria Zitz}
\authornote{Both authors contributed equally to this research.}
\email{valeria.zitz@kit.edu}
\orcid{0009-0004-1158-861X}
\affiliation{%
  \institution{Karlsruhe Institute of Technology}
  \city{Karlsruhe}
  \country{Germany}
}

\author{Michael K\"{u}ttner}
\authornotemark[1]
\email{michael.kuettner@kit.edu}
\orcid{0009-0000-9021-0359}
\affiliation{%
  \institution{Karlsruhe Institute of Technology}
  \city{Karlsruhe}
  \country{Germany}
}

\author{Jonas Hummel}
%\authornotemark[1]
\email{jonas.hummel@kit.edu}
\orcid{0009-0005-8563-6175}
\affiliation{%
  \institution{Karlsruhe Institute of Technology}
  \city{Karlsruhe}
  \country{Germany}
}

\author{Taku Inoue}
%\authornotemark[1]
\email{inoue.taku.18@shizuoka.ac.jp}
\orcid{0009-0004-3421-5939}
\affiliation{%
  \institution{Shizuoka University}
  \city{Hamamatsu}
  \country{Japan}
}

\author{Tobias R\"{o}ddiger}
\email{tobias.roeddiger@ipai-foundation.ai}
\orcid{0000-0002-4718-9280}
\affiliation{%
  \institution{IPAI Foundation gGmbH}
  \city{Heilbronn}
  %\state{}
  \country{Germany}
}

\author{Michael Beigl}
\email{michael.beigl@kit.edu}
\orcid{0000-0001-5009-2327}
\affiliation{%
  \institution{Karlsruhe Institute of Technology}
  \city{Karlsruhe}
  \country{Germany}
}

%%
%% By default, the full list of authors will be used in the page
%% headers. Often, this list is too long, and will overlap
%% other information printed in the page headers. This command allows
%% the author to define a more concise list
%% of authors' names for this purpose.
\renewcommand{\shortauthors}{Valeria Zitz et al.}
%% No italics, no superscripts, not anonymous
%% Use footnote or author note to identify equal contribution, shared contribution, and/or contact author info

%%
%% The abstract is a short summary of the work to be presented in the
%% article.

%Wir zeigen, wie ihr runter kommt, HR geht runter und ganz konkret sagen was wir zeigen können 
\begin{abstract}
  We present \systemname{}, a walk-up demo of a closed-loop infrared earable that uses smart-ring heart rate to create subtle, ear-localized warming cues. Building on work on thermal comfort and in-ear infrared stimulation, \systemname{} explores warmth at the ear as a
  biosignal-adaptive %body-based
  cue for brief recovery moments following acute activation. A smartphone first establishes an individual resting baseline and derives a personalized heart-rate threshold, while the earable controller independently enforces an over-temperature cut-off and communication fail-safes. During the guided demo flow, attendees complete a brief rapid-breathing activation phase until their heart rate reaches the personalized threshold. This triggers an ear-localized warming cue followed by a short relaxation phase, while physiological changes are displayed on a live dashboard. Attendees can also manually explore different infrared stimulation intensities. To contextualize the demo, we report preliminary placebo-controlled UX ratings from 18 participants: participants rated the active prototype higher on perceived relaxation and perceived recovery support than an identical-looking placebo. Together, the demo illustrates how infrared earables can make physiological feedback tangible through subtle, biosignal-adaptive
  %body-based
  thermal cues.
\end{abstract}

%%
%% The code below is generated by the tool at http://dl.acm.org/ccs.cfm.
%% Please copy and paste the code instead of the example below.
%%


%%
%% Keywords. The author(s) should pick words that accurately describe
%% the work being presented. Separate the keywords with commas.
%% A "teaser" image appears between the author and affiliation
%% information and the body of the document, and typically spans the
%% page.
\begin{teaserfigure}
  \includegraphics[width=\textwidth]{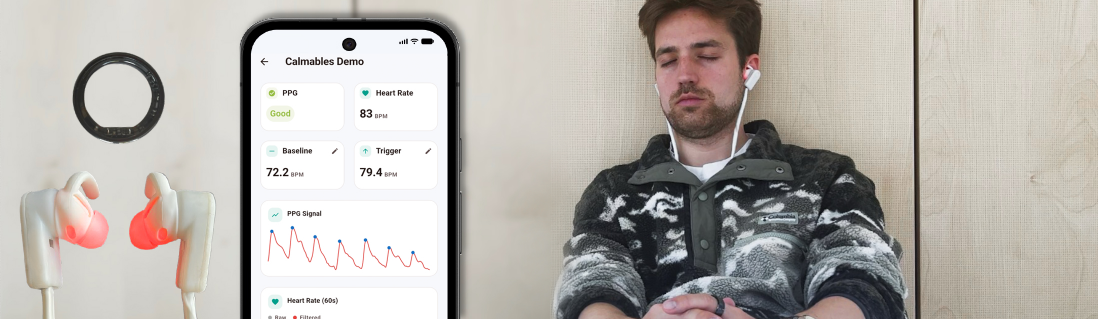}
  \caption{\systemname{} system overview. A smart ring streams PPG data via BLE to a smartphone, which estimates heart rate and executes the closed-loop control policy. The smartphone then sends BLE commands to the earable controller to actuate infrared stimulation. The Dashboard mode shown in the figure displays the live PPG signal, estimated heart rate, baseline, trigger threshold, and current stimulation state, allowing attendees to inspect and compare manual and HR-responsive control.
  %overview. A smart ring streams PPG data via BLE to a smartphone, which estimates heart rate and runs the closed-loop policy. BLE commands are sent to an earable controller that actuates infrared stimulation. The smartphone shows Dashboard mode, in which attendees can compare manual and HR-responsive control while observing live heart rate, baseline, and stimulation state. %; the guided demo flow is shown in \autoref{fig:screens}.
  %tr{die erwähnung von safety limits etc. finde ich irgendwie komisch, der fokus sollte auf der experience und der anwendung sein, auch die details mit BLE sind vielleicht zu technisch, würde schauen, dass es primär um die anwendung geht}
  }
  \Description{Composite overview of the Calmables system. On the left, a smart ring is shown above two white infrared earpieces glowing red. In the center, a smartphone dashboard displays PPG quality, a heart rate of 83 BPM, a baseline of 72.2 BPM, a trigger threshold of 79.4 BPM, and a live PPG waveform. On the right, a seated person wears the earable with eyes closed and hands resting in their lap.}
  \label{fig:teaser}
\end{teaserfigure}

%\received{30 May 2025}

%%
%% This command processes the author and affiliation and title
%% information and builds the first part of the formatted document.
\maketitle
\newpage
\section{Introduction}

Acute stress is common in everyday work and study routines, while recovery opportunities are often brief and constrained by ongoing obligations~\cite{sonnentag2022recovery}. Established relaxation practices can be effective, but often require dedicated time, sustained attention, or a supportive environment~\cite{grossman2004mindfulness,khoury2015mindfulness,manzoni2008relaxation}. This motivates subtle 
biosignal-adaptive %body-based 
cues that can be experienced during short recovery moments without relying on screens or audio.

Warmth is a promising cue for such interactions: prior work links thermal stimulation and thermal comfort to stress-related subjective states such as anxiety, distress, and discomfort \cite{kennedy2023randomized,lin2025effects}. Yet, most work applies warmth to body regions other than the ear or focuses on thermal comfort rather than acute stress recovery. Recent thermal ear-worn systems include \emph{Warmth on Demand} for heated headphones and \emph{Heatables}, which showed that in-ear infrared stimulation can influence thermal perception and comfort~\cite{knierim2024, zitz2025}.

Building on this work, we present \systemname{}, an experimental demo that applies closed-loop infrared ear warming to an acute stress-recovery scenario. Attendees experience biosignal-adaptive feedback as active, ear-localized warmth, compare manual and heart rate (HR)-responsive modes, and inspect how HR changes relative to a resting baseline activate thermal stimulation. \systemname{} explores whether active infrared ear stimulation changes the subjective recovery experience and can be delivered in a perceivable, comfortable, and interpretable way.

\section{Related Work}
\label{sec:rw}

\paragraph{Subtle Wearable Cues for Recovery}

Short pauses during work can provide opportunities for recovery, but interactive support for such moments must remain lightweight, unobtrusive, and easy to understand~\cite{albulescu2022give,hunter2016give,kim2022,kim2018job}. Wearable biofeedback systems have explored low-effort cues for down-regulation and emotion regulation, often using vibrotactile feedback, rhythmic guidance, or interoceptive feedback~\cite{Yu2018,SCHOELLER2019310,costa2016emotioncheck,costa2019boostmeup,choi2020ambienbeat}. While biosignal-based stress detection is well established, lightweight closed-loop systems that translate physiological signals into subtle interventions remain comparatively underexplored~\cite{giannakakis_review_2022,taskasaplidis_review_2024,jimenez-ocana_systematic_2023}. In this demo, we therefore use heart rate not as a specific stress detector, but as an inspectable control signal for baseline-relative thermal feedback.

Prior work also shows that stimulation design choices, including body location, can shape comfort, perceptibility, and perceived restfulness~\cite{jueun2025,valente2024modulating}. Compared to haptic, visual, or audio feedback, thermal cues remain less explored in wearable biofeedback, despite offering a private, low-attention modality that can be experienced without screens or sound. \systemname{} builds on this opportunity by making thermal feedback directly perceivable and inspectable in a short walk-up interaction.

\paragraph{Thermal Cues and Actuation}

Warmth is relevant for recovery-oriented interaction because thermal stimulation and thermal comfort have been linked to autonomic and subjective state regulation. Skin heating and cooling can modulate autonomic nervous activity~\cite{Kinugasa1999SkinCoolingHeatingAutonomic}, cutaneous temperature manipulation has been shown to influence sleep-related depth~\cite{Raymann2008SkinDeep}. Infrared or photobiomodulation interventions have reported effects primarily in subjective relaxation, sleep quality, or well-being-related outcomes~\cite{kennedy2023randomized,lin2025effects}. These findings motivate warmth as a plausible 
biosignal-adaptive %body-based 
cue, but they do not directly establish ear-localized warming as an acute stress-recovery intervention.

Ear-worn devices provide an interesting platform for such cues because they are socially familiar, close to physiological signals, and compatible with everyday wearable form factors~\cite{roeddiger2022}. Recent systems have begun to explore thermal actuation around or inside the ear. \emph{Warmth on Demand} investigated heated headphones as a personal thermal comfort system, while \emph{Heatables} demonstrated that in-ear infrared stimulation can influence thermal perception and comfort~\cite{zitz2025}. Building on this work, our demo applies ear-localized infrared warming beyond thermal comfort and explores it as a subtle cue during acute stress recovery. We frame this as an experimental User Experience (UX) exploration: attendees can experience active warming, compare manual and HR-responsive behavior, and inspect how heart rate triggers thermal stimulation.

%Unlike prior systems that rely on custom ultrasonic hardware or directional phased arrays, our approach uses a minimal infrastructure: a single ultrasonic source and an off-the-shelf wearable earable device. We demonstrate that commodity tweeters and passive ultrasonic playback can be leveraged for multi-user, localized, and multilingual audio experiences in public spaces—without requiring beamforming, pairing, or active playback control.

\section{\systemnameheading{}: Closed-Loop Infrared Earables}
\label{sec:system}

\systemname{} reuses the in-ear IR/NIR-LED actuator introduced with \emph{Heatables}~\cite{zitz2025}. We replaced its wired controller with a Bluetooth-enabled microcontroller and integrated the earable into the OpenWearable app. New to \systemname{} are the BLE communication, smart-ring PPG/IMU processing, signal-quality-aware HR estimation, baseline-relative control, dashboard, and guided demo flow. Thus, Heatables provides the actuator, while this work contributes the sensing, closed-loop control, and interactive mechanisms.

%\systemname{} consists of a smart-ring, an Android phone, and an infrared earable based on the \emph{Heatables} module~\cite{zitz2025}. The system is designed for a short experience in which users can feel ear-localized warming and inspect how heart rate changes are mapped to bounded thermal actuation. Sensing and visualization run on the phone, while the earable enforces actuation and safety locally.

%\begin{figure*}[!t]
 %   \centering
 %   \includegraphics[width=\linewidth,trim=0cm 0cm 0cm 0.5cm]{figures/Demo-Pipeline2.pdf}
 %   \caption{Data and control flow of \systemname{}. The smart-ring streams PPG/IMU data to the phone, which estimates heart rate and signal quality, maintains a session baseline, and sends bounded BLE commands in manual or HR-responsive mode. The earable drives the infrared LEDs, reports telemetry, and enforces local safety limits such as temperature, duty cycle, and fail-safe shutoff.}
 %   \label{fig:chain}
%\end{figure*}

\begin{figure*}[!t]
    \centering
    \includegraphics[width=\linewidth,trim=0cm 0cm 0cm 0.5cm]{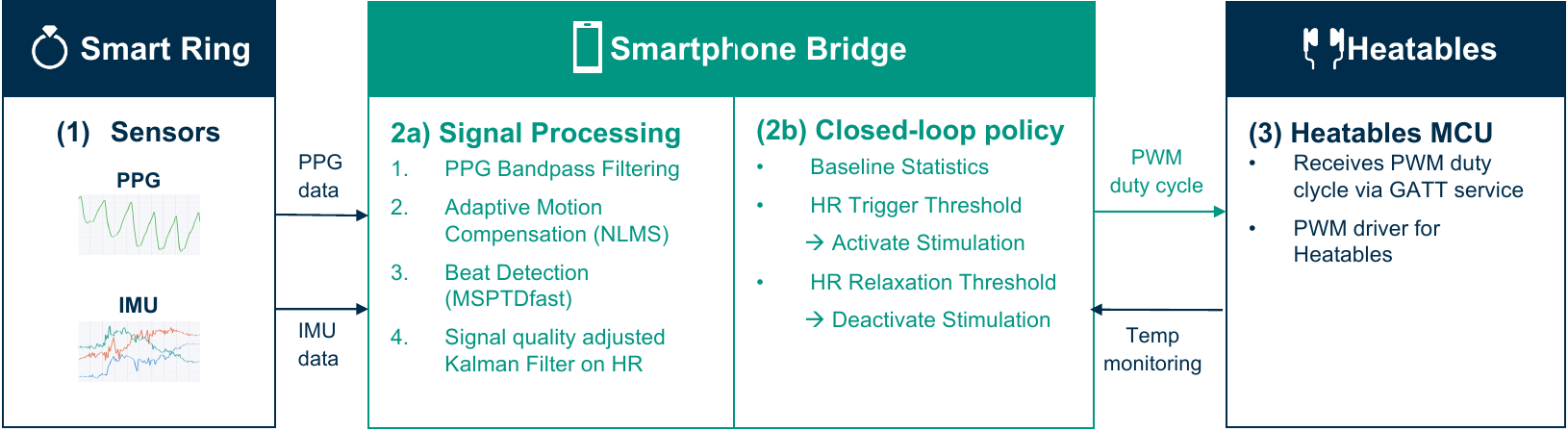}
    \caption{Data and control flow of \systemname{}. The smart-ring streams PPG/IMU data to the phone, which estimates HR and signal quality, maintains a session baseline, and sends BLE commands in manual or HR-responsive mode. The earable drives the infrared LEDs, and returns the device temperature while safely handling faults such as connection loss.}
    \Description{Block diagram with three stages arranged from left to right. The smart ring provides PPG and IMU sensor data to the smartphone bridge. The smartphone first performs band-pass filtering, motion compensation, beat detection, and signal-quality-adjusted heart-rate filtering, then applies a closed-loop policy using baseline statistics and heart-rate thresholds to activate or deactivate stimulation. It sends a PWM duty cycle to the Heatables microcontroller, which drives the infrared earpieces, while temperature-monitoring data flows back to the smartphone.}
    \label{fig:chain}
\end{figure*}

\subsection{Data and Control Flow}
\label{subsec:data-control-flow}

The ring streams photoplethysmography (PPG) and inertial measurement unit (IMU) data to the smartphone, which estimates HR and signal quality. The earable uses pulse-width modulation (PWM)-controlled infrared LEDs for fine-grained adjustment of the warming intensity, with the aggregate electrical power across both earpieces limited to approximately $650\,\mathrm{mW}$. A temperature sensor placed at the multi-chip emitter monitors the housing temperature and immediately disables stimulation at $50\,^{\circ}\mathrm{C}$; BLE loss or controller faults likewise trigger shutdown.

The app establishes a session baseline either through a short seated calibration phase or manual entry. During calibration, the user remains calm and relaxed while the baseline HR is estimated over a $30\,\mathrm{s}$ window. The trigger threshold is defined as $\theta = \mu_{\mathrm{HR}} + 3\sigma_{\mathrm{HR}}$, where $\mu_{\mathrm{HR}}$ and $\sigma_{\mathrm{HR}}$ denote the mean and standard deviation of the calibration window.

In HR-responsive mode, stimulation is activated at the selected PWM intensity after the smoothed HR exceeds $\theta$ for three consecutive samples. It is deactivated after HR remains below the hysteresis threshold $\theta_{\mathrm{off}} = \mu_{\mathrm{HR}} + 0.2(\theta-\mu_{\mathrm{HR}})$ for three samples. Manual mode allows direct PWM control, enabling attendees to experience the infrared cue even when their HR remains near baseline. The data and control flow are summarized in \autoref{fig:chain}.

%The dashboard visualizes signal quality, heart rate, baseline, trigger threshold, current PWM value, and controller state. This makes the control behavior inspectable during the demo and supports quick switching between manual and HR-responsive modes. 

\subsection{Heart Rate Assessment}

%We employ a multistage pipeline for robust heart rate estimation, illustrated in \autoref{fig:chain}. 
For heartbeat detection, the green PPG channel is used, as it is generally less susceptible to motion artifacts. The signal is first band-pass filtered between $0.5\,\mathrm{Hz}$ and $8\,\mathrm{Hz}$ to suppress baseline drift and high-frequency noise.

To further reduce motion artifacts, adaptive motion compensation is performed using the ring's IMU data. The gravity component is estimated via an exponential moving average (EMA) and removed from the accelerometer signal. The resulting motion signals are normalized per axis and used as references for a normalized least mean squares (NLMS) adaptive filter to estimate and suppress motion-related distortions in the PPG signal.

Heartbeat detection is performed using MSPTDfast~\cite{charlton2025msptdfast} on sliding windows of $8\,\mathrm{s}$ with a stride of $1\,\mathrm{s}$. 
From the detected PPG pulse peaks, we derive inter-beat intervals (IBIs), i.e., the time differences between consecutive detected pulse beats. To suppress implausible detections and stabilize the estimated HR, a Kalman filter is applied to the IBI-derived HR estimates. %To suppress implausible beat detections and stabilize the estimated HR,  a Kalman filter is applied to HR estimates derived from the detected RR intervals. 
The HR dynamics are modeled as a random walk with latent state $x_k$ and observed HR $z_k$:
\[
x_k = x_{k-1} + w_k, \qquad w_k \sim \mathcal{N}(0, P_k)
\]
\[
z_k = x_k + v_k, \qquad v_k \sim \mathcal{N}(0, R_k)
\]
The measurement noise variance $R_k$ is adapted based on the estimated PPG signal quality, which is determined from motion intensity, peak amplitude variability, and IBI stability. We classify the signal quality $Q(t)$ into four discrete states: \textsc{good}, \textsc{fair}, \textsc{bad}, and \textsc{unavailable}. The corresponding $R_k$ values were experimentally determined through comparison of the measured HR with a reference ECG measurement.
Based on the estimated quality, the measurement noise variance is adjusted as
\[
R_k =
\begin{cases}
0.9, & Q(t)=\textsc{good} \\
4.8, & Q(t)=\textsc{fair} \\
\infty, & Q(t)\in\{\textsc{bad}, \textsc{unavailable}\}.
\end{cases}
\]
For low-quality signals, HR updates are therefore either strongly smoothed or completely suppressed.

\subsection{Preliminary UX Comparison}

\begin{figure}[b]
    \centering
    \includegraphics[width=\linewidth]{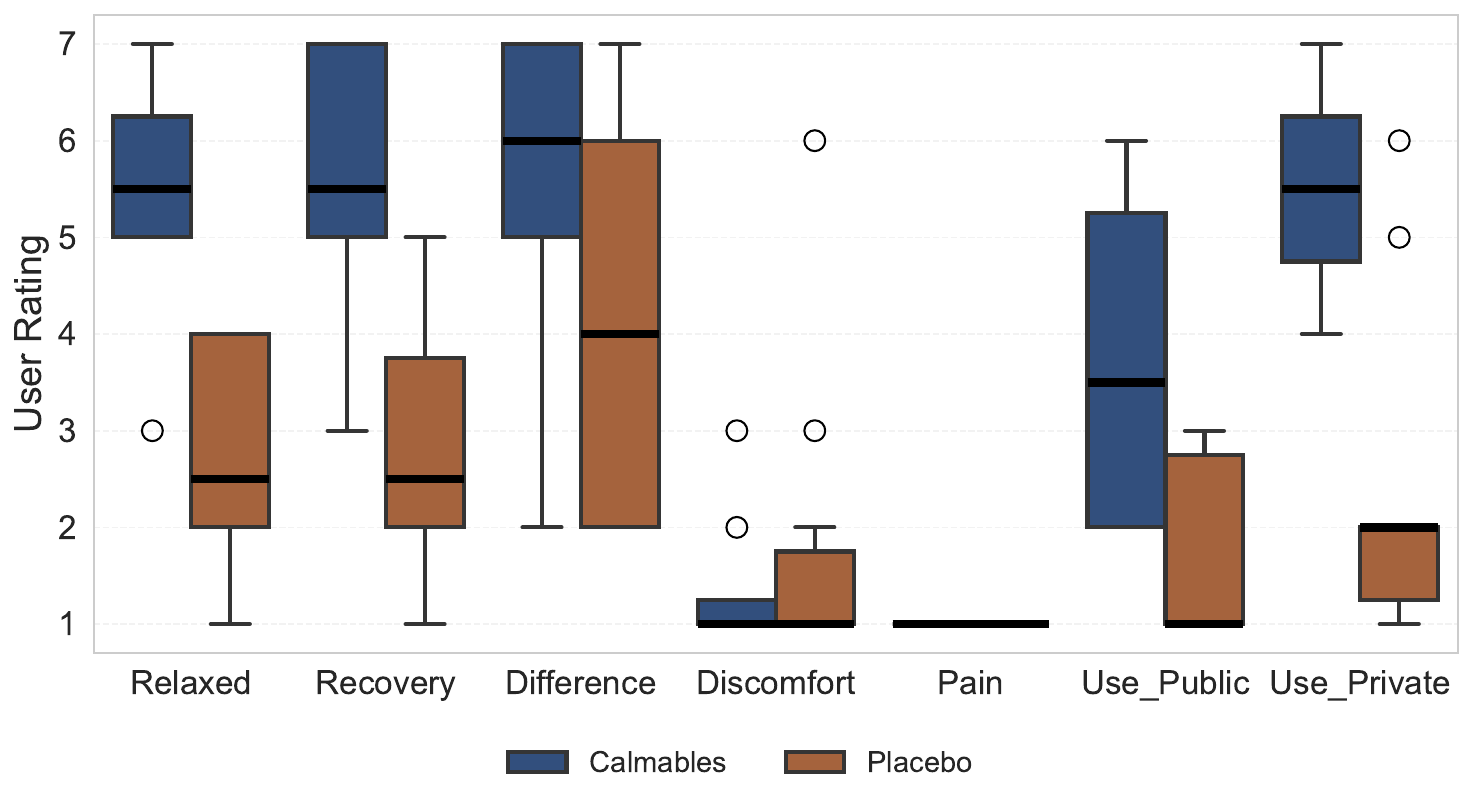}
    \caption{Preliminary UX ratings after the recovery scenario following acute stress induction for the active infrared prototype and a visually similar placebo. Higher values indicate stronger agreement with each item.}
    \Description{Grouped boxplots compare Calmables and placebo ratings on a scale from 1 to 7 across seven questionnaire items. Calmables has higher median ratings for relaxation, recovery support, difference from regular earbuds, intended public use, and intended private use. Both conditions have low discomfort ratings and the minimum pain rating. The largest visible separations between conditions occur for relaxation, recovery support, and intended private use.}
    \label{fig:UX_ratings}
\end{figure}

We conducted an initial evaluation of Calmables with \DemoN{} participants (\DemoGenderFemaleN{} female; 8 male;
$M_\text{age} = \DemoAgeMean{}$). Following stress induction via the Maastricht Acute Stress Test (MAST)~\cite{shilton_maastricht_2017}, participants wore either Calmables or a visually similar placebo version without infrared stimulation during a 15-minute recovery period. Afterwards, they completed a questionnaire assessing their user experience (Likert-scale items with $1$--$7$). All comparisons used bootstrapped \textit{t}-tests ($N_\text{boot} = 10,000$) to account for the small sample size.
Consistent with our directional hypotheses, we used one-sided bootstrapped tests~\cite{cho_is_2013} and applied Bonferroni correction for multiple comparisons ($\alpha_\text{Bonf} = 0.05 / \UxNTests{} = \UxAlphaBonf{}$). Results are depicted in \autoref{fig:UX_ratings}.

\begin{figure*}[!t]
    \centering
    \includegraphics[width=\linewidth,trim=0cm 0cm 0cm 0.5cm]{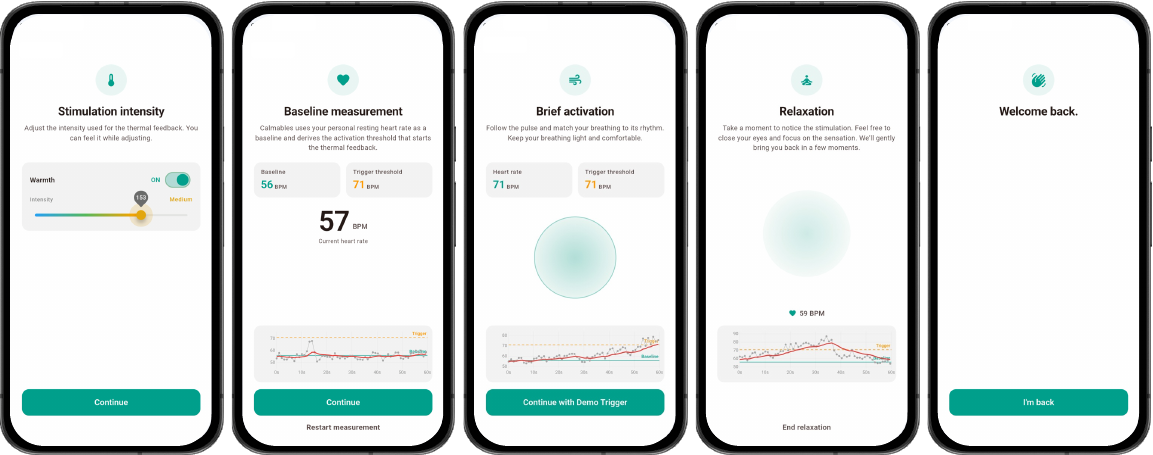}
    \caption{Screens of the \systemname{} demo flow: users first select a preferred stimulation intensity, followed by resting baseline measurement and a brief activation phase. Once heart rate crosses the personalized trigger threshold, ear-localized warming is activated to support relaxation. As heart rate returns toward baseline, stimulation is automatically deactivated and the user is guided back to the final screen. Live PPG and heart-rate visualizations make the sensing-to-actuation loop transparent throughout the interaction.
    %Screens of the \systemname{} demo flow: selection of the preferred stimulation intensity, baseline measurement, activation phase, automatic warming for relaxation after HR crosses the personalized trigger, returning screen with automatic deactivation as HR returns toward baseline. Live PPG and HR visualizations make the sensing-to-actuation loop inspectable throughout the flow.
    }
    \Description{Five smartphone screens show the guided demo sequence from left to right. The first screen provides an on-off control and slider for selecting stimulation intensity. The second measures a resting baseline and displays the current heart rate, baseline, trigger threshold, and a live graph. The third guides a brief breathing activation while showing heart rate relative to the trigger. The fourth presents the relaxation phase with a soft circular visual, current heart rate, and a graph of heart rate returning toward baseline. The fifth welcomes the user back and provides a button to finish the flow.}
    \label{fig:screens}
\end{figure*}

Compared to the placebo, participants rated the active prototype higher on perceived relaxation ($M_\text{C} = \UxRelaxedMeanCalmables$, $M_\text{P} =
\UxRelaxedMeanPlacebo$; $p = \UxRelaxedPOne$) and perceived recovery support ($M_\text{C} = \UxRecoveryMeanCalmables$,
$M_\text{P} = \UxRecoveryMeanPlacebo$; $p = \UxRecoveryPOne$). Participants perceived both devices as distinct from wearing regular earbuds ($M_\text{C} =
\UxDifferenceMeanCalmables$, $M_\text{P} = \UxDifferenceMeanPlacebo$, $p = \UxDifferencePOne$). Neither device caused pain
($M_\text{C} = \UxPainMeanCalmables$, $M_\text{P} = \UxPainMeanPlacebo$)
or notable discomfort ($M_\text{C} = \UxDiscomfortMeanCalmables$,
$M_\text{P} = \UxDiscomfortMeanPlacebo$; $p =
\UxDiscomfortPOne$), indicating good physical acceptance for both devices. Regarding future use intent, participants expressed a significantly higher preference for using Calmables in private settings ($M_\text{C} =
\UxUsePrivateMeanCalmables$, $M_\text{P} = \UxUsePrivateMeanPlacebo$;
$p = \UxUsePrivatePOne$), whereas intent to use it in public was numerically higher but did not pass the corrected significance threshold ($M_\text{C} = \UxUsePublicMeanCalmables$, $M_\text{P} = \UxUsePublicMeanPlacebo$; $p = \UxUsePublicPOne$).

\section{Demonstration}
\label{sec:demo}

We demonstrate \systemname{} through an interactive walk-up setup comprising the BLE-controlled Heatables, a smart ring, and an Android phone (\autoref{fig:screens}). Two complementary modes let attendees experience ear-localized warming and inspect how HR drives thermal actuation.

%\begin{figure}[t]
%    \centering
%    \includegraphics[width=\linewidth]{figures/demo-platzhalter.pdf}
%    \caption{Demo setup: a user wears the infrared earable and smart-ring while the smartphone dashboard visualizes HR, session baseline, stimulation events, earable temperature, and controller state in real time.}
 %   \label{fig:demo}
%\end{figure}

In \emph{Dashboard mode}, attendees inspect live HR, baseline, trigger threshold, and stimulation level. They can explore infrared intensities using manual PWM control or activate HR-responsive control, which starts warming when HR exceeds the personalized trigger. This mode supports open-ended comparison of manual and automatic control.

The \emph{Guided Live Demo} provides a time-condensed recovery experience. Participants first preview the warming cue and select a comfortable intensity. A 30-second baseline measurement then establishes the baseline HR and trigger threshold as described in \autoref{subsec:data-control-flow}. A pulsing visual subsequently guides increasingly fast breathing while live HR and the thresholds remain visible. The breathing task provides a brief, equipment-free activation, as changes in breathing pattern and lung volume can elicit rapid transient HR responses~\cite{mehlsen1987heart}. Warming starts when the smoothed HR exceeds the trigger for three consecutive samples. If this does not occur within 20 seconds, an optional demo trigger starts the same actuation path without altering or simulating HR.

Participants are then invited to breathe normally and notice the warmth. The relaxation phase lasts at least 25 seconds, after which warming ends when HR remains below the hysteresis threshold for three samples; a manual end becomes available after 30 seconds. Repeated phone vibration marks completion until the participant confirms their return. Two one-tap questions capture perceived warmth and relaxation, followed by a summary of the baseline, peak HR, trigger source, intensity, and ratings. Completed runs can be reviewed when leaving the flow.

Together, the modes combine open exploration with a repeatable end-to-end experience. HR serves as an inspectable control signal rather than a reliable stress classifier, while all stimulation uses the same BLE control path and device-side safety enforcement.
%Together, these options allow attendees to inspect the full interaction loop: how a seated baseline, live HR changes, control mode, and stimulation limits shape the delivered warming cue. The demo therefore foregrounds the experience and interpretability of HR-responsive thermal feedback rather than stress classification.
%Attendees can compare baseline-dependent triggering, HR-responsive versus manual control, and optional activation as a way to make the mapping visible without treating it as stress classification. By the end of the demo, attendees have experienced infrared earables as a private, low-attention cue channel and have inspected how a seated resting baseline can serve as an adaptive reference for bounded thermal stimulation.

\section{Conclusion}
\label{sec:conclusion}

We presented \systemname{}, a closed-loop infrared earable demo for subtle, 
biosignal-adaptive %body-based 
biofeedback in short recovery moments. The system combines smart-ring heart rate sensing, smartphone-based signal processing and visualization, as well as safe infrared-based warming stimulation through the BLE-controlled Heatables. Through our interactive walk-up demo, attendees can experience ear-localized warming, compare manual and HR-responsive modes, and inspect how baseline-relative heart rate triggers activate thermal stimulation. Our preliminary placebo-controlled UX comparison suggests that active infrared stimulation changes the subjective recovery experience while remaining comfortable. %however, we interpret these findings as early UX evidence rather than as physiological relaxation or stress-recovery effects. 
%We hope this 
%demonstration 
This motivates future placebo- and treatment-controlled studies with larger and more diverse samples, combining subjective stress and recovery items with physiological markers such as HRV, respiration, salivary cortisol, and alpha-amylase, and contributes to the development of safe, interpretable thermal feedback technologies for everyday recovery scenarios.

\begin{acks}
This work was partially funded by the ZEISS Innovation Hub and the Deutsche Forschungsgemeinschaft (DFG, German Research Foundation) -- GRK2739/2 -- Project Nr. 447089431 -- Research Training Group: KD2School -- Designing Biosignal-Adaptive Systems for Decision-Making Processes
\end{acks}

%\newpage

%%
%% The next two lines define the bibliography style to be used, and
%% the bibliography file.
\balance
\bibliographystyle{ACM-Reference-Format}
\bibliography{references}

%%
%% If your work has an appendix, this is the place to put it.

\end{document}

%% file: statistics.tex
\newcommand{\DemoN}{18}

\newcommand{\DemoAgeMean}{25.28}

\newcommand{\DemoGenderFemaleN}{10}

\newcommand{\UxNTests}{7}
\newcommand{\UxAlphaBonf}{0.0071}

\newcommand{\UxRelaxedMeanCalmables}{5.50}
\newcommand{\UxRelaxedMeanPlacebo}{2.80}

\newcommand{\UxRelaxedPOne}{0.0004}

\newcommand{\UxRecoveryMeanCalmables}{5.62}
\newcommand{\UxRecoveryMeanPlacebo}{2.90}

\newcommand{\UxRecoveryPOne}{0.0013}

\newcommand{\UxDifferenceMeanCalmables}{5.62}
\newcommand{\UxDifferenceMeanPlacebo}{4.10}

\newcommand{\UxDifferencePOne}{0.0532}

\newcommand{\UxDiscomfortMeanCalmables}{1.38}
\newcommand{\UxDiscomfortMeanPlacebo}{1.80}

\newcommand{\UxDiscomfortPOne}{0.2484}

\newcommand{\UxPainMeanCalmables}{1.00}
\newcommand{\UxPainMeanPlacebo}{1.00}

\newcommand{\UxUsePublicMeanCalmables}{3.75}
\newcommand{\UxUsePublicMeanPlacebo}{1.70}

\newcommand{\UxUsePublicPOne}{0.0106}

\newcommand{\UxUsePrivateMeanCalmables}{5.50}
\newcommand{\UxUsePrivateMeanPlacebo}{2.40}

\newcommand{\UxUsePrivatePOne}{0.0018}

%% file: references.bib
@inproceedings{costa2016emotioncheck,
  title={EmotionCheck: leveraging bodily signals and false feedback to regulate our emotions},
  author={Costa, Jean and Adams, Alexander T and Jung, Malte F and Guimbreti{\`e}re, Fran{\c{c}}ois and Choudhury, Tanzeem},
  booktitle={Proceedings of the 2016 ACM international joint conference on pervasive and ubiquitous computing},
  pages={758--769},
  year={2016}
}

@article{costa2019boostmeup,
  title={Boostmeup: Improving cognitive performance in the moment by unobtrusively regulating emotions with a smartwatch},
  author={Costa, Jean and Guimbreti{\`e}re, Fran{\c{c}}ois and Jung, Malte F and Choudhury, Tanzeem},
  journal={Proceedings of the ACM on Interactive, Mobile, Wearable and Ubiquitous Technologies},
  volume={3},
  number={2},
  pages={1--23},
  year={2019},
  publisher={ACM New York, NY, USA}
}

@inproceedings{choi2020ambienbeat,
  title={ambienBeat: Wrist-worn mobile tactile biofeedback for heart rate rhythmic regulation},
  author={Choi, Kyung Yun and Ishii, Hiroshi},
  booktitle={Proceedings of the fourteenth international conference on tangible, embedded, and embodied interaction},
  pages={17--30},
  year={2020}
}

@inproceedings{jueun2025,
author = {Lee, Jueun and Filpe, Martin and Lepold, Philipp and R\"{o}ddiger, Tobias and Beigl, Michael},
title = {Haptic Biofeedback for Wakeful Rest: Does Stimulation Location Make a Difference?},
year = {2025},
isbn = {9798400714818},
publisher = {Association for Computing Machinery},
address = {New York, NY, USA},
url = {https://doi.org/10.1145/3715071.3750427},
doi = {10.1145/3715071.3750427},
booktitle = {Proceedings of the 2025 ACM International Symposium on Wearable Computers},
pages = {83–90},
numpages = {8},
location = {Espoo, Finland},
series = {ISWC '25}
}

@inproceedings{valente2024modulating,
  title={Modulating heart activity and task performance using haptic heartbeat feedback: A study across four body placements},
  author={Valente, Andreia and Lee, Dajin and Choi, Seungmoon and Billinghurst, Mark and Esteves, Augusto},
  booktitle={Proceedings of the 37th Annual ACM Symposium on User Interface Software and Technology},
  pages={1--13},
  year={2024}
}

@article{kennedy2023randomized,
  title={A randomized, sham-controlled trial of a novel near-infrared phototherapy device on sleep and daytime function},
  author={Kennedy, Kathryn ER and Wills, Chloe CA and Holt, Catie and Grandner, Michael A},
  journal={Journal of Clinical Sleep Medicine},
  volume={19},
  number={9},
  pages={1669--1675},
  year={2023},
  publisher={American Academy of Sleep Medicine}
}

@article{lin2025effects,
  title={Effects of photobiomodulation on the sleep quality and quality of life of night-shift nurses},
  author={Lin, Mei-Yen and Chen, Chia-Chi and Chen, Ming-Jie and Hwang, Lih-Lian and Wu, Jih-Huah and Su, Chuan-Tsung},
  journal={Lasers in Medical Science},
  volume={40},
  number={1},
  pages={221},
  year={2025},
  publisher={Springer}
}

@inproceedings{zitz2025,
author = {Zitz, Valeria and K\"{u}ttner, Michael and Hummel, Jonas and Knierim, Michael T. and Beigl, Michael and R\"{o}ddiger, Tobias},
title = {Heatables: Effects of Infrared-LED-Induced Ear Heating on Thermal Perception, Comfort, and Cognitive Performance},
year = {2025},
isbn = {9798400714818},
publisher = {Association for Computing Machinery},
address = {New York, NY, USA},
url = {https://doi.org/10.1145/3715071.3750421},
doi = {10.1145/3715071.3750421},
booktitle = {Proceedings of the 2025 ACM International Symposium on Wearable Computers},
pages = {91–97},
numpages = {7},
location = {Espoo, Finland},
series = {ISWC '25}
}

@ARTICLE{Yu2018,
    
AUTHOR={Yu, Bin  and Funk, Mathias  and Hu, Jun  and Wang, Qi  and Feijs, Loe },
           
TITLE={Biofeedback for Everyday Stress Management: A Systematic Review},
          
JOURNAL={Frontiers in ICT},
          
VOLUME={Volume 5 - 2018},
  
YEAR={2018},
  
URL={https://www.frontiersin.org/journals/ict/articles/10.3389/fict.2018.00023},
  
DOI={10.3389/fict.2018.00023},
  
ISSN={2297-198X}}

@article{SCHOELLER2019310,
title = {Enhancing human emotions with interoceptive technologies},
journal = {Physics of Life Reviews},
volume = {31},
pages = {310-319},
year = {2019},
note = {Physics of Mind},
issn = {1571-0645},
doi = {https://doi.org/10.1016/j.plrev.2019.10.008},
url = {https://www.sciencedirect.com/science/article/pii/S1571064519301599},
author = {F. Schoeller and A.J.H. Haar and A. Jain and P. Maes}
}

@article{Raymann2008SkinDeep,
  author  = {Raymann, R. J. and Swaab, D. F. and Van Someren, E. J.},
  title   = {Skin Deep: Enhanced Sleep Depth by Cutaneous Temperature Manipulation},
  journal = {Brain: A Journal of Neurology},
  year    = {2008},
  volume  = {131},
  number  = {Pt 2},
  pages   = {500--513},
  doi     = {10.1093/brain/awm315}
}

@article{Kinugasa1999SkinCoolingHeatingAutonomic,
  author  = {Kinugasa, H. and Hirayanagi, K.},
  title   = {Effects of Skin Surface Cooling and Heating on Autonomic Nervous Activity and Baroreflex Sensitivity in Humans},
  journal = {Experimental Physiology},
  year    = {1999},
  volume  = {84},
  number  = {2},
  pages   = {369--377}
}

@article{albulescu2022give,
  title={"Give me a break!" A systematic review and meta-analysis on the efficacy of micro-breaks for increasing well-being and performance},
  author={Albulescu, Patricia and Macsinga, Irina and Rusu, Andrei and Sulea, Coralia and Bodnaru, Alexandra and Tulbure, Bogdan Tudor},
  journal={PloS one},
  volume={17},
  number={8},
  pages={e0272460},
  year={2022},
  publisher={Public Library of Science}
}

@article{hunter2016give,
  title={Give me a better break: Choosing workday break activities to maximize resource recovery.},
  author={Hunter, Emily M and Wu, Cindy},
  journal={Journal of Applied Psychology},
  volume={101},
  number={2},
  pages={302},
  year={2016},
  publisher={American Psychological Association}
}

@article{kim2022,
  title={Daily microbreaks in a self-regulatory resources lens: Perceived health climate as a contextual moderator via microbreak autonomy.},
  author={Kim, Sooyeol and Cho, Seonghee and Park, YoungAh},
  journal={Journal of Applied Psychology},
  volume={107},
  number={1},
  pages={60},
  year={2022},
  publisher={American Psychological Association}
}

@article{kim2018job,
  title={Daily micro-breaks and job performance: General work engagement as a cross-level moderator.},
  author={Kim, Sooyeol and Park, YoungAh and Headrick, Lucille},
  journal={Journal of Applied Psychology},
  volume={103},
  number={7},
  pages={772},
  year={2018},
  publisher={American Psychological Association}
}

@article{shilton_maastricht_2017,
	title = {The {Maastricht} {Acute} {Stress} {Test} ({MAST}): {Physiological} and {Subjective} {Responses} in {Anticipation}, and {Post}-stress},
	volume = {8},
	issn = {1664-1078},
	shorttitle = {The {Maastricht} {Acute} {Stress} {Test} ({MAST})},
	url = {https://www.frontiersin.org/journals/psychology/articles/10.3389/fpsyg.2017.00567/full},
	doi = {10.3389/fpsyg.2017.00567},
	language = {English},
	urldate = {2026-04-02},
	journal = {Frontiers in Psychology},
	publisher = {Frontiers},
	author = {Shilton, Alexandra L. and Laycock, Robin and Crewther, Sheila G.},
	month = apr,
	year = {2017},
}

@article{cho_is_2013,
	series = {Advancing {Research} {Methods} in {Marketing}},
	title = {Is two-tailed testing for directional research hypotheses tests legitimate?},
	volume = {66},
	issn = {0148-2963},
	url = {https://www.sciencedirect.com/science/article/pii/S0148296312000550},
	doi = {10.1016/j.jbusres.2012.02.023},
	number = {9},
	urldate = {2025-04-28},
	journal = {Journal of Business Research},
	author = {Cho, Hyun-Chul and Abe, Shuzo},
	month = sep,
	year = {2013},
	pages = {1261--1266},
}

@inproceedings{knierim2024,
author = {Knierim, Michael Thomas and Braun, Lukas and Perusquia-Hernandez, Monica},
title = {Warmth on Demand: Designing Headphones for Enhanced Thermal Comfort in Work Environments},
year = {2024},
isbn = {9798400703317},
publisher = {Association for Computing Machinery},
address = {New York, NY, USA},
url = {https://doi.org/10.1145/3613905.3650884},
doi = {10.1145/3613905.3650884},
booktitle = {Extended Abstracts of the CHI Conference on Human Factors in Computing Systems},
articleno = {369},
numpages = {7},
location = {Honolulu, HI, USA},
series = {CHI EA '24}
}

@article{roeddiger2022,
author = {R\"{o}ddiger, Tobias and Clarke, Christopher and Breitling, Paula and Schneegans, Tim and Zhao, Haibin and Gellersen, Hans and Beigl, Michael},
title = {Sensing with Earables: A Systematic Literature Review and Taxonomy of Phenomena},
year = {2022},
issue_date = {September 2022},
publisher = {Association for Computing Machinery},
address = {New York, NY, USA},
volume = {6},
number = {3},
url = {https://doi.org/10.1145/3550314},
doi = {10.1145/3550314},
journal = {Proc. ACM Interact. Mob. Wearable Ubiquitous Technol.},
month = sep,
articleno = {135},
numpages = {57}
}

@article{sonnentag2022recovery,
  title={Recovery from work: Advancing the field toward the future},
  author={Sonnentag, Sabine and Cheng, Bonnie Hayden and Parker, Stacey L},
  journal={Annual Review of Organizational Psychology and Organizational Behavior},
  volume={9},
  number={1},
  pages={33--60},
  year={2022},
  publisher={Annual Reviews}
}

@article{khoury2015mindfulness,
  author  = {Khoury, Bassam and Sharma, Manoj and Rush, Sarah E. and Fournier, Claude},
  title   = {Mindfulness-based stress reduction for healthy individuals: A meta-analysis},
  journal = {Journal of Psychosomatic Research},
  year    = {2015},
  volume  = {78},
  number  = {6},
  pages   = {519--528},
  doi     = {10.1016/j.jpsychores.2015.03.009}
}

@article{grossman2004mindfulness,
  author  = {Grossman, Paul and Niemann, Ludger and Schmidt, Stefan and Walach, Harald},
  title   = {Mindfulness-based stress reduction and health benefits: A meta-analysis},
  journal = {Journal of Psychosomatic Research},
  year    = {2004},
  volume  = {57},
  number  = {1},
  pages   = {35--43},
  doi     = {10.1016/S0022-3999(03)00573-7}
}

@article{manzoni2008relaxation,
  author  = {Manzoni, Gian Mauro and Pagnini, Francesco and Castelnuovo, Gianluca and Molinari, Enrico},
  title   = {Relaxation training for anxiety: A ten-years systematic review with meta-analysis},
  journal = {BMC Psychiatry},
  year    = {2008},
  volume  = {8},
  pages   = {41},
  doi     = {10.1186/1471-244X-8-41}
}

@article{giannakakis_review_2022,
	title = {Review on {Psychological} {Stress} {Detection} {Using} {Biosignals}},
	volume = {13},
	issn = {1949-3045, 2371-9850},
	url = {https://ieeexplore.ieee.org/document/8758154/},
	doi = {10.1109/TAFFC.2019.2927337},
	language = {en},
	number = {1},
	urldate = {2026-04-13},
	journal = {IEEE Transactions on Affective Computing},
	author = {Giannakakis, Giorgos and Grigoriadis, Dimitris and Giannakaki, Katerina and Simantiraki, Olympia and Roniotis, Alexandros and Tsiknakis, Manolis},
	month = jan,
	year = {2022},
	pages = {440--460},
}

@article{taskasaplidis_review_2024,
	title = {Review of {Stress} {Detection} {Methods} {Using} {Wearable} {Sensors}},
	volume = {12},
	issn = {2169-3536},
	url = {https://ieeexplore.ieee.org/abstract/document/10458924},
	doi = {10.1109/ACCESS.2024.3373010},
	urldate = {2026-04-13},
	journal = {IEEE Access},
	author = {Taskasaplidis, Georgios and Fotiadis, Dimitris A. and Bamidis, Panagiotis D.},
	year = {2024},
	pages = {38219--38246},
}

@article{jimenez-ocana_systematic_2023,
	title = {A {Systematic} {Review} of {Technology}-{Aided} {Stress} {Management} {Systems}: {Automatic} {Measurement}, {Detection} and {Control}},
	volume = {11},
	issn = {2169-3536},
	url = {https://ieeexplore.ieee.org/document/10287332/},
	doi = {10.1109/ACCESS.2023.3325763},
	urldate = {2026-04-13},
	journal = {IEEE Access},
	author = {Jiménez-Ocaña, Alvaro A. and Pantoja, Andrés and Valderrama, Mario Andrés and Giraldo, Luis Felipe},
	year = {2023},
	pages = {116109--116126},
}

@article{charlton2025msptdfast,
  title={The MSPTDfast photoplethysmography beat detection algorithm: design, benchmarking, and open-source distribution},
  author={Charlton, Peter H and Arg{\"u}ello-Prada, Erick Javier and Mant, Jonathan and Kyriacou, Panicos A},
  journal={Physiological Measurement},
  volume={46},
  number={3},
  pages={035002},
  year={2025},
  publisher={IOP Publishing}
}

@article{mehlsen1987heart,
  author  = {Mehlsen, J. and Pagh, K. and Nielsen, J. S. and Sestoft, L. and Nielsen, S. L.},
  title   = {Heart Rate Response to Breathing: Dependency upon Breathing Pattern},
  journal = {Clinical Physiology},
  year    = {1987},
  volume  = {7},
  number  = {2},
  pages   = {115--124},
  doi     = {10.1111/j.1475-097X.1987.tb00153.x}
}
